\documentclass[preprint,amsmath,amssymb]{revtex4}

\usepackage{graphicx}% Include figure files
\usepackage{bm}% bold math

\newcommand{\Puw}{P_{\mu w}}
\newcommand{\Pint}{P_\mathrm{int}}

\begin{document}
\title{A semi-empirical model for two-level system noise in
superconducting microresonators }
\begin{abstract}
\normalsize{ We present measurements of the low--temperature excess
frequency noise of four niobium superconducting coplanar waveguide
microresonators, with center strip widths $s_r$ ranging from
3~$\mu$m to 20~$\mu$m. For a fixed internal power, we find that the
frequency noise decreases rapidly with increasing center strip
width, scaling as $1/s_r^{1.6}$. We show that this geometrical
scaling is readily explained by a simple semi-empirical model which
assumes a surface distribution of independent two-level system
fluctuators. These results allow the resonator geometry to be
optimized for minimum noise.  }
\end{abstract}
\author{Jiansong Gao}
\affiliation{Division of Physics, Mathematics, and Astronomy,
California Institute of Technology, Pasadena, CA 91125}

\author{Miguel Daal}%
\affiliation{ Physics Department, University of California at
Berkeley, Berkeley, CA 94720}

\author{Bernard Sadoulet}%
\affiliation{ Physics Department, University of California at
Berkeley, Berkeley, CA 94720}

\author{Benjamin A. Mazin}
\affiliation{Jet Propulsion Laboratory, California Institute of
Technology, Pasadena, CA 91109}

\author{Peter K. Day}
\affiliation{Jet Propulsion Laboratory, California Institute of
Technology, Pasadena, CA 91109}

\author{Henry G. Leduc}
\affiliation{Jet Propulsion Laboratory, California Institute of
Technology, Pasadena, CA 91109}

\author {Anastasios Vayonakis}
\affiliation{Division of Physics, Mathematics, and Astronomy,
California Institute of Technology, Pasadena, CA 91125}

\author{Jonas Zmuidzinas}
\affiliation{Division of Physics, Mathematics, and Astronomy,
California Institute of Technology, Pasadena, CA 91125}

\author{John M. Martinis}
\affiliation{Department of Physics, University of California, Santa
Barbara, California 93106}
\date{\today}% It is always \today, today,
       % but any date may be explicitly specifie

% 77.22.-d Dielectric properties of solids and liquids
% 85.25.Oj Superconducting optical, X-ray, and gamma-ray detectors (SIS, NIS, transition edge)
% 72.70.+m Noise processes and phenomena
% 61.43.Fs Glasses
\maketitle

%\small
%1. introduction
%intro to superconducting microwave resonators
%intro to excess frequency noise
%intro to paper A
%contents of this paper
Thin-film superconducting microresonators are of great interest for
a number of applications (see
[1-4]\nocite{Mazin02,Day03,Kumar08,Gao08a} and references therein).
Excess frequency noise is universally observed in these
resonators\cite{Day03,Mates08,Baselmans08} and is very likely caused
by two-level systems (TLS) in dielectric
materials\cite{Gao07,Kumar08}. Indeed, the TLS hypothesis is
supported by the observed dependence of the noise on resonator
internal power\cite{Gao07,Gao06} and temperature.\cite{Kumar08} In a
recent paper\cite{Gao08a} (Paper A hereafter), we presented
measurements of the TLS--induced low--temperature frequency shifts
of five Niobium ($T_c = 9.2$~K) coplanar waveguide (CPW) resonators
with varying center strip widths $s_r$. From the observed
geometrical scaling of the frequency shifts ($\sim 1/s_r$), we
showed that the TLS must be located in a thin (few nm) layer on the
surface of the CPW. In this letter, we propose a semi-empirical TLS
noise model that assumes this surface distribution, and we show that
the model explains our measurements of the geometrical scaling of
the noise.

%2. device and experimental setup
%description of the device
%noise measurement setup
%noise data analysis
The device used for the experiment in this paper is exactly the same
device as used for Paper A. In brief, the chip contains
five CPW quarter-wavelength resonators ($Z_0 \approx 50~\Omega,
f_r\approx 6$~GHz) made by patterning a 120~nm-thick Nb film
deposited on a crystalline sapphire substrate.
Each resonator is
capacitively coupled to a common feedline, using a CPW coupler
($Q_c\sim 50,000$) of length $l_c\cong200~\mu$m and with a common
center-strip width of $s_c=3~\mu$m. The coupler is then widened into
the resonator body, with a center-strip width of $s_r=$ 3, 5, 10, 20
or 50~$\mu$m, and a length of $l_r\sim5~$mm. The noise was measured
using a standard $IQ$ homodyne technique\cite{Day03,Kumar08};
both the measurement setup and the analysis of the noise
data are identical to our previous work\cite{Gao07}.

%3. basic results
%noise spectra at fixed readout power
%noise v.s. internal power
%noise v.s. center strip width (uncorrected)
%figure12

The device is cooled in a dilution refrigerator to a base
temperature of 55~mK. The fractional frequency noise spectra
$S_{\delta f}(\nu)/f_r^2$ of the five resonators were measured for
microwave readout power $\Puw$ in the range $-61$~dBm to $-73$~dBm;
the $-65$~dBm spectra are shown in Fig.~1(a). We clearly see that
the noise has a common spectral shape but decreases as the center
strip becomes wider. Unfortunately, the data for the lowest--noise
($50~\mu$m) resonator are influenced by the noise floor of our
cryogenic microwave amplifier, so we exclude this resonator from
further discussion. The noise levels at $\nu=$2~kHz were retrieved
from the noise spectra and are plotted as a function of resonator
internal power $\Pint = 2Q_r^2 \Puw/\pi Q_c$ in
%Fig.~\ref{fig:noise_p}.
Fig.~1(b). All resonators display a power dependence close to
$S_{\delta f}/f_r^2 \propto P_\mathrm{int}^{-1/2}$ as we have
previously observed\cite{Gao06,Gao07,Kumar08}. In order to study the
geometrical scaling of the noise in more detail, we first fit the
noise vs. power data for each resonator to a simple power law, and
retrieve the values of the noise $S_{\delta
f}(2~\mathrm{kHz})/f_r^2$ at $\Pint=-25$~dBm for each geometry.
These results (Fig.~2) again show that the noise decreases with
increasing $s_r$, although not (yet) as a simple power law.

To make further progress, we introduce a semi-empirical
model for the TLS noise. We assume that the TLS have a uniform
spatial distribution within a volume of TLS--hosting material $V_h$
that occupies some portion of the total resonator volume $V$.
Consider a TLS labeled $\alpha$, located at a random position $\vec
r_\alpha \in V_h$ and with an energy level separation $E_\alpha
=(\Delta_\alpha^2 + \Delta_{0,\alpha}^2)^{1/2}$.
%Here we are using
%the standard notation\cite{Phillips87} for the TLS asymmetry
%$\Delta_\alpha$ and tunnel splitting $\Delta_{0,\alpha}$. As usual,
%these are assumed to be random and to have uniform and log--uniform
%distributions, respectively; their joint distribution function is
Here $\Delta_\alpha$ and $\Delta_{0,\alpha}$ are the TLS asymmetry
energy and tunnel splitting, which are random and have a joint distribution function
$f(\Delta, \Delta_0) = P / \Delta_0$ as introduced by
Phillips\cite{Phillips72}. The TLS transition dipole moment is given
by $\vec d_\alpha = \hat{n}_\alpha d_0 \Delta_{0,\alpha} /
E_\alpha$, where $d_0$ is the maximum dipole moment for a TLS with
energy $E_\alpha$ and the dipole orientation unit vector
$\hat{n}_\alpha$ is assumed to be random and isotropically
distributed. In the weak--field, linear response limit,
the TLS contribution to the dielectric tensor of the hosting medium is
%\begin{eqnarray}
%&&\Delta \epsilon_{kl}(\omega, \vec r) = -\sum_\alpha d_{\alpha, k}
%d_{\alpha, l} \delta(\vec r - \vec r_\alpha)\nonumber\\
%&&\times \left[
%  \frac{1}{(E_\alpha - \hbar \omega) + j\Gamma_\alpha} +
%  \frac{1}{(E_\alpha + \hbar \omega) - j\Gamma_\alpha}
%\right] \, \sigma_{z,\alpha} \label{eqn:TLSsum}
%\end{eqnarray}
\begin{equation}
\Delta \epsilon_{kl}(\omega, \vec r) = -\sum_\alpha d_{\alpha, k}
d_{\alpha, l} \delta(\vec r - \vec r_\alpha)
\chi_\alpha(\omega)
\sigma_{z,\alpha} \label{eqn:TLSsum}
\end{equation}
where $k, l$ represent Cartesian components,
$\chi_\alpha(\omega) = 1/(E_\alpha - \hbar \omega + j \Gamma_\alpha)
                     + 1/(E_\alpha + \hbar \omega - j \Gamma_\alpha)$
is a damped single-pole response function for $e^{+j \omega t}$
harmonic time dependence,
and $\sigma_{z,\alpha}$ is the usual diagonal Pauli operator that takes
values of $-1$ for the lower state of the TLS and $+1$ for the upper
state. Averaging over the TLS position, asymmetry, tunnel splitting,
and dipole orientation, and assuming a thermal distribution for the
level population, the TLS contribution to the
(isotropic) dielectric function is given by
\begin{eqnarray}
&&\left< \Delta \epsilon(\omega) \right> = \int_0^{E_\mathrm{max}}
 \frac{P d_0^2}{3}\tanh\left( \frac{E}{2 k_B
T}\right)\chi(\omega) dE
\nonumber\\
%&&\times\left[
%  \frac{1}{(E - \hbar \omega) + j \Gamma} +
%  \frac{1}{(E + \hbar \omega) - j \Gamma}
%\right] \,\nonumber\\
&&= -\frac{2 P d_0^2}{3\epsilon}\left[\Psi \left( \frac{1}{2} -
\frac{\hbar \omega-j\Gamma}{2 j \pi k_B T }\right) - \log
\frac{E_\mathrm{max}}{2\pi k_B T}\right]\label{eqn:depsave}
\end{eqnarray}
where $\chi(\omega) = 1/(E - \hbar \omega + j \Gamma) + 1/(E + \hbar
\omega - j \Gamma)$,~$E_\mathrm{max}$ is the maximum energy level
separation, and $\Psi$ is the complex digamma function.
%For $\Gamma << E$, the TLS contribution to the dielectric loss
%tangent $\delta$ may be simplified to
%\begin{equation}
%\tand(\omega, T) = - \frac{\im \left< \Delta \epsilon(\omega) \right>}{\epsilon}
%      = \frac{\pi  P d_0^2(\hbar \omega)}{3 \epsilon} \,
%         \tanh\left( \frac{\hbar \omega}{2 k_B T}\right)\ ,
%\end{equation}
%which can be used to express the fractional
%perturbation of the reactive (real) part of the dielectric constant
%as
%
%\begin{eqnarray}
%\frac{\re \left< \Delta \epsilon(\omega) \right>}{\epsilon}
% &=& - \frac{2 \tand(\omega, 0)}{\pi}\nonumber\\
%&\times&
% \left[
%\re \Psi \left( \frac{1}{2} + \frac{1}{2 \pi j} \frac{\hbar
%\omega}{k_B T}\right) - \log \frac{E_\mathrm{max}}{2\pi k_B T}
%\right]\ .
%\end{eqnarray}
%assuming constant $\delta_\mathrm{TLS}(\omega,0)$. These are
%well-known results\cite{Phillips87, Hunklinger76}, and allow the
%average fractional frequency shift of a resonator to be computed
%using\cite{Gao08a, Pozar}
The real ($\Delta \epsilon_1)$ and imaginary ($\Delta \epsilon_2)$
parts of Eq.~(\ref{eqn:depsave}) yield the
well--known results for the TLS contribution to
the dielectric constant\cite{Hunklinger76} and
loss tangent\cite{Phillips72, Martinis05}.
The former allows the temperature--dependent
fractional  frequency shift of a resonator to be computed
using\cite{Gao08a}
\begin{equation}
\frac{\left< \Delta f_r \right>}{f_r} = - \frac{\int_{V_h}
\left< \Delta \epsilon_1 \right> |\vec E|^2\, d\vec r}
     {2\int_{V} \epsilon |\vec E|^2\, d\vec r}\ .
\end{equation}

This result provides an excellent description of the experimental
data\cite{Kumar08,Gao08a} at $T<<T_c$.

Now, if the dielectric constant fluctuates on time scales
$\tau_\epsilon \gg 1/\omega$, we would expect to see resonator
frequency fluctuations given by
\begin{equation}
\frac{\delta f_r(t) }{f_r} = - \frac{\int_{V_h}
\delta \epsilon_1(\vec r, t) |\vec E|^2\, d\vec r}
     {2\int_{V} \epsilon |\vec E|^2\, d\vec r}\ .
\end{equation}
From Eq.~(\ref{eqn:TLSsum}), we see that $\Delta \epsilon_1$ could
fluctuate with time if the TLS switch states randomly
($\sigma_{z,\alpha}$ changes sign), for instance due to phonon
emission or absorption, or if the the energy level separation
$E_\alpha$ is perturbed randomly, for instance due to a collection
of nearby TLS that randomly switch states and produce a
randomly--varying strain field that couples to TLS $\alpha$.
Whatever the mechanism, for independently fluctuating TLS, from
Eq.~(\ref{eqn:TLSsum}) we would expect that the Fourier spectra of
the $\delta \epsilon_1$ fluctuations to obey $\left<\delta
\epsilon_1^*(\vec r_1, \nu_1)\,\delta \epsilon_1(\vec r_2, \nu_2)
\right > = S_\epsilon(\vec r_1, \nu_1, T) \delta(\vec r_1 - \vec
r_2) \delta(\nu_1 - \nu_2)$.
%\begin{equation}
%\left<\delta \epsilon_1^*(\vec r_1, \nu_2)\,\delta \epsilon_1(\vec
%r_2, \nu_2) \right > = S_\epsilon(\vec r_1, \nu_1, T) \delta(\vec
%r_1 - \vec r_2) \delta(\nu_1 - \nu_2)
%\end{equation}
Therefore, the resonator frequency power spectrum should be given by
\begin{equation}
\frac{S_{\delta f_r}(\nu)}{f_r^2} =  \frac{\int_{V_h}
S_\epsilon(\vec r, \nu, T) |\vec E|^4 d \vec r} {4\left(\int_V
\epsilon |\vec{E}|^2 d \vec r\right)^2}\ . \label{eqn:Sdf3}
\end{equation}
If $S_\epsilon$ is independent of the field strength $|\vec E|$,
Eq.~(\ref{eqn:Sdf3}) predicts that the resonator noise is
independent of microwave power, contrary to our
observations\cite{Gao07, Gao06,Kumar08} which are made at the
relatively high power levels of interest for detector applications.
As we have argued previously\cite{Gao07}, TLS saturation effects are
very likely responsible for the observed power dependence of the
noise. The saturation of TLS dissipation is a well known
effect;\cite{Hunklinger76,Phillips87,Martinis05} we therefore make
the ansatz that the noise depends on field strength in a similar
manner:
%\begin{eqnarray}
%S_\epsilon(\vec r, \nu, T) =
%\frac{\kappa(\nu,T)}{\sqrt{|\vec{E}(\vec r)|^2 + E_{n,c}^2}}
%%\kappa(\nu,T)/\sqrt{|\vec{E}(\vec r)|^2 + E_{n,c}^2}
%\label{eqn:kappa3}
%\end{eqnarray}
\begin{equation}
S_\epsilon(\vec r, \nu, \omega, T) = \kappa(\nu,\omega,T)/\sqrt{|\vec{E}(\vec r)|^2
+ E_{n,c}^2(\omega,T)} \, ,\label{eqn:kappa3}
\end{equation}
where $E_{n,c}(\omega,T)$ is a critical electric field,
likely related to the critical field for the saturation of the TLS dissipation,
and the noise spectral density coefficient $\kappa(\nu, \omega, T)$ is allowed to
vary with (microwave) frequency $\omega$ and temperature\cite{Kumar08}.
Because we
are assuming a uniform distribution of TLS in the volume $V_h$, we
do not expect $S_\epsilon$ to have an additional explicit dependence
on position $\vec r$. At high power for which $E \gg E_{n,c}$ in
the region contributing significantly to the resonator noise,
Eq.~(\ref{eqn:Sdf3}) becomes
\begin{eqnarray}
\frac{S_{\delta f_r}(\nu)}{f_r^2} = \kappa(\nu, \omega, T)\, \frac{
\,\int_{V_h}|\vec E|^3 d^3r}{4\left(\int_{V} \epsilon |\vec{E}|^2
d^3r\right)^2}\label{eqn:dfreskappa3}
\end{eqnarray}
which exhibits the desired $\Pint^{-1/2}$ scaling with power.
%5. noise scaling
%E^3 weighting
%correction formula, noise v.s. center strip width (corrected), s^{-1.58}

Eq.~(\ref{eqn:dfreskappa3}) implies that the noise contributions are
weighted by $|\vec{E}|^3$, so TLS fluctuators located near the
coupler end of a quarter-wave resonator should give significantly
larger noise contributions than those located near the shorted end.
Therefore, for the resonators that are wider than the coupler
($s_r\neq s_c=3~\mu$m), the measured values of $S_{\delta
f_r}/f_r^2$ need to be corrected for the coupler's noise
contribution. A similar procedure was applied in Paper A to correct
the frequency shift data. In the limit $l_c << l_r$, the correction
is given by $S_{\delta f_r}^* = ( S_{\delta f_r} - \eta S_{\delta
f_r,~3\mu \textrm{m}})/(1-\eta)$, where $\eta =3\pi l_c/4(l_c+l_r)$.
The corrected values are plotted in Fig.~2 and are found to have a
simple power--law scaling $1/s_r^{1.58}$. We find a similar noise
scaling, $1/s_r^{\alpha}$ with $\alpha$ between 1.49 and 1.6, for
noise frequencies $400~\mathrm{Hz} <\nu<3~\mathrm{kHz}$.
%figure2

%6. E field from conformal mapping
%zero thickness, s^{-1} for volume distribution and s^{-2} for surface distribution, favor a surface distribution
%mapping with edge slope
%result of E^3 v.s. t0 of a CPW with center strip normalized to 1\mu m, assuming different edge slope, fit to t^\gamma
%E^3(s-t) = E^3(t0=t/s)/s^2 which scales as s^{2-\gamma}, 2-\gamma ~ -1.55 agrees with -1.58
%mention E^3 from volume distribution scales as ?, so rule out

While the fact that an $|E|^3$--weighted coupler noise correction
leads to a simple power law noise scaling is already quite
encouraging, we will now go further and show that the observed
$s_r^{-1.58}$ power--law slope can be reproduced by our model.
Measurements of the anomalous low-temperature frequency shift
described in Paper A have already pointed to a surface distribution
of TLS. If these TLS are also responsible for the frequency noise,
according to Eq.~(\ref{eqn:dfreskappa3}) we would expect the noise
to have the same geometrical scaling as the contour integral $I_3
=\int |\vec{E}|^3 ds$ evaluated either on the metal surface
($I^m_3$) or the exposed substrate surface ($I^g_3$). For
zero-thickness CPW, although the integral is divergent, the expected
scaling can be shown to be $I_3\propto  1/s_r^2$. For CPW with
finite thickness, we can evaluate $I_3$ numerically using the
electric field derived from a numerical conformal mapping solution.
The two-step mapping procedure used here is modified from that given
by Collin\cite{Collin} and is illustrated in the inset of Fig.~3. We
first map a quadrant of finite-thickness CPW with half thickness $t$
( in the $W$-plane) to a zero-thickness CPW (in the $Z$-plane) and
then to a parallel-plate capacitor (in the $\xi$-plane).
%The mapping parameters are found with the help of Matlab
%Schwarz-Christoffel Toolbox\cite{SCtoolbox}.
To avoid non-integrable singularities, we must constrain all
internal angles on the conductor edges to be less than $\pi/2$,
which leads to the condition $0.25 <\beta < 0.5$, where $\beta\pi$
is the angle defined in Fig.~3.

Instead of evaluating $I_3$ directly, we define a normalized
dimensionless integral $F_3(t, s_r) =\int |\vec E/E^*|^3 ds^*$,
where $s^* = s/s_r$ is a normalized integration coordinate and $E^*
= V/s_r$  is a characteristic field strength for a CPW with voltage
$V$. Now $F_3$ depends only on the ratio $t/s_r$ and is related to
the original contour integral by $I_3(s_r,t,V) = (V^3/s_r^2)
F_3(t/s_r)$. The results $F^m_3(t/s_r)$ calculated for the metal
surface are plotted in Fig.~3, and show a power law scaling
$F^m_3\sim (t/s_r)^\gamma$ with $\gamma \approx -0.45$ for
$0.003<t/s_r<0.02$, the relevant range for our experiment. We also
find that for a wide range of $\beta$, $0.27<\beta<0.43$, although
the absolute values of $F^m_3(t/s_r)$ vary significantly, the
scaling index $\gamma$ remains almost constant,
$-0.456<\gamma<-0.440$. Therefore, $\gamma$ appears to depend little
on the edge shape.

From Eq.~(\ref{eqn:dfreskappa3}), the noise scaling is predicted to
be $I^m_3(t, s_r, V) \propto s_r^{-2-\gamma} \sim s_r^{-1.55}$ (at
fixed $V$), which agrees surprisingly well with the measured
$s_r^{-1.58}$ scaling. We also investigated the case for TLS located
on the exposed substrate surface, and found that $F^g_3$ has almost
identical scaling ($\gamma\approx-0.45$) as $F^m_3$. While still
cannot say whether the TLS are on the surface of the metal or the
exposed substrate, we can safely rule out a volume distribution of
TLS fluctuators in the bulk substrate; this assumption yields a
noise scaling of $\sim s_r^{-1.03}$, significantly different than
measured.
%figure3

%7. conclusions
%1. surface TLS fluctators
%2. non-oxidizing metal, hybrid geometry
%3. if field and TLS both known, can determine k. then with k, should predict noise in
%other resonator geometries, type. In our case, t, edge shape unkown,
%scaling doesn't depend on t, edge if they are common, so still
%useful in predict noise on other resonator on the same wafer at least.
In summary, the scaling of the frequency noise
with resonator power and CPW geometry can be satisfactorily
explained by a semi-empirical model assuming a surface distribution
of independent TLS fluctuators. These results allow the
resonator geometry to be optimized.
For example, one can design a quarter-wave CPW kinetic
inductance detector\cite{Mazin02,Day03} which is wider
on the coupler end to benefit from the noise reduction, but narrower
at the low-$|\vec E|$ shorted end to maintain a high kinetic
inductance fraction and responsivity.
If the spatial distribution of the TLS and the $\vec{E}$ field are both known,
values of $\kappa$ and $E_{n,c}$ can be determined, allowing noise
predictions to be made using Eq.~(\ref{eqn:dfreskappa3}).
Unfortunately we do not know the exact $\vec{E}$ field distribution
for our CPW resonators because of the sensitivity to the edge shape,
nor do we know the thickness of the TLS surface layer.
Future experiments with simplified geometries and
at lower powers should allow our ansatz (Eq.~(\ref{eqn:kappa3})) to
be tested, and may yield a quantitative determination of
$\kappa(\nu,\omega,T)$.

We thank Clare Yu and Sunil Golwala for useful discussions. The
device was fabricated in the University of California, Berkeley,
Microfabrication Laboratory. This work was supported in part by the
NASA, NSF, JPL, and the Gordon and Betty Moore Foundation.

\newpage

%\bibliography{noise}% Produces the bibliography via BibTeX.

\newpage
\begin{center}
FIGURE CAPTIONS
\end{center}
\vspace{1cm}

\noindent
Figure 1\\
(Color online) Frequency noise of the four CPW resonators measured
at $T=$55~mK. (a) Frequency noise spectra at $\Puw=$-65~dBm. From
top to bottom, the four curves correspond to CPW center strip widths
of $s_r$ = 3~$\mu$m, 5~$\mu$m, 10~$\mu$m, and 20~$\mu$m. The various
spikes seen in the spectra are due to pickup of stray signals by the
electronics and cabling. (b) Frequency noise at $\nu=$2~kHz as a
function of $P_\mathrm{int}$. The markers represent different
resonator geometries, as indicated by the values of $s_r$ in the
legend. The dashed lines indicate power law fits to the data of each
geometry.
\vspace{0.5cm}

\noindent
Figure 2\\
(Color online) The measured frequency noise $S_{\delta
f}(2~\mathrm{kHz})/f_r^2$ at $\Pint=-25$~dBm is plotted as a
function of the center strip width $s_r$. Values directly retrieved
from power-law fits to the data in Fig.~1 are indicated by the open
squares. Values corrected for the coupler's contribution are
indicated  by the stars. The corrected values of $S_{\delta
f}(2~\mathrm{kHz})/f_r^2$ scale as $s_r^{-1.58}$, as indicated by
the dashed line.
\vspace{0.5cm}

\noindent
Figure 3\\
(Color online) The calculated dimensionless
noise scaling function $F^m_3(t/s_r)$ is plotted as a function of
the ratio between the CPW half film thickness $t$ and the center
strip width $s_r$. The inset shows the conformal mapping used to
derive the electric field. The contour integral for $F^m_3(t/s_r)$
is evaluated on the surface of the metal, as outlined by the solid
lines in the $W$-plane. Results are shown for four different values
of the parameter $\beta = 0.28,~0.33,~0.38,~0.43$ that controls the
edge shape (see inset). The dashed lines indicate power law
$(t/s_r)^\gamma$ fits to $F^m_3(t/s_r)$.

\newcommand{\placefigures}{
\newpage
\begin{figure}
  \begin{center}
  \resizebox{6.5in}{!}{\includegraphics{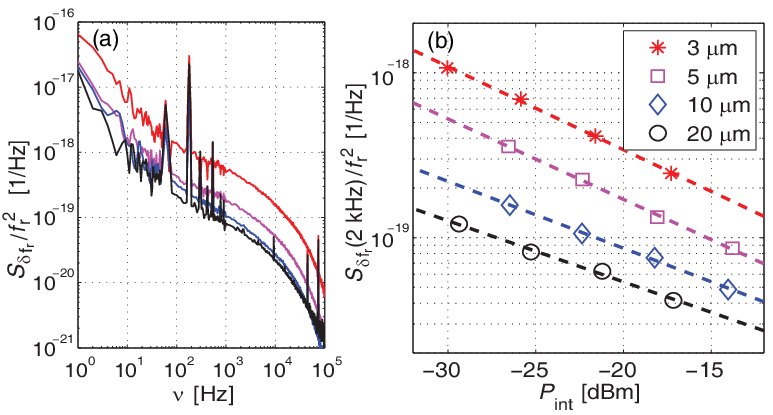}}\\
\vspace{1in} Figure 1
  \end{center}
\end{figure}
\newpage
\begin{figure}
  \begin{center}
  \resizebox{6.5in}{!}{\includegraphics{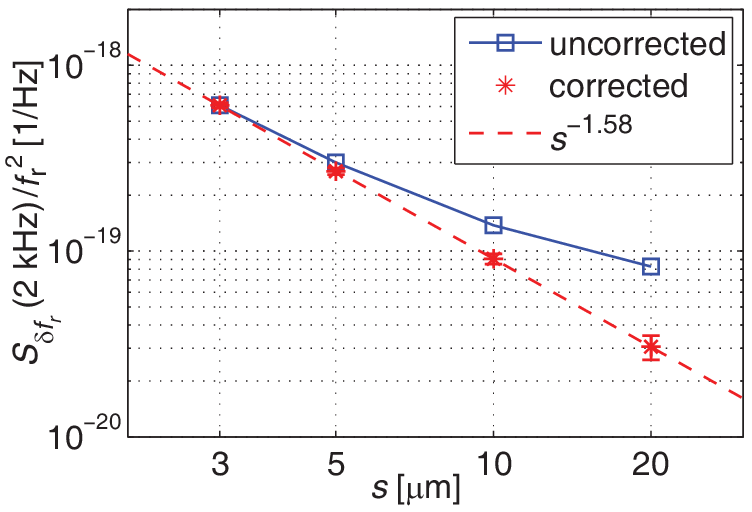}}\\
\vspace{1in} Figure 2
  \end{center}
\end{figure}
\newpage
\begin{figure}
  \begin{center}
  \resizebox{6.5in}{!}{\includegraphics{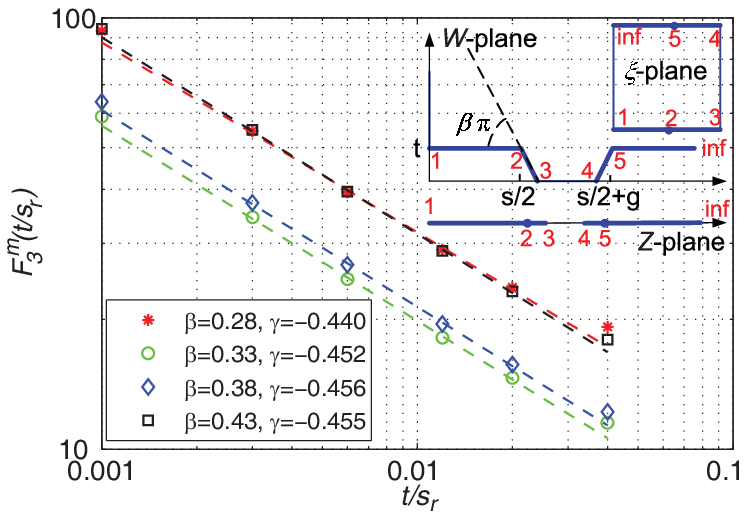}}\\
\vspace{1in} Figure 3
  \end{center}
\end{figure}
}

\placefigures
\end{document}